\documentclass[aps,preprint]{revtex4}
\usepackage{amsfonts}
\usepackage{amsmath}
\usepackage{amssymb}
\usepackage{mathrsfs}
\usepackage{subfigure}
\usepackage{graphicx}
\usepackage{epstopdf}
\usepackage{float}
\usepackage{color}
\usepackage{bm}
\usepackage{ulem}
\usepackage{xcolor}
\usepackage{appendix}

\begin{document}
\title{Effects of Born-Infeld electrodynamics on black hole shadows}
\preprint{CTP-SCU/2022006}
\author{Aoyun He}
\email{heaoyun@stu.scu.edu.cn} 
\author{Jun Tao}
\email{taojun@scu.edu.cn}
\author{Peng Wang}
\email{pengw@scu.edu.cn}
\author{Yadong Xue}
\email{xueyadong@stu.scu.edu.cn}
\author{Lingkai Zhang}
\email{zhanglingkai@stu.scu.edu.cn}
\affiliation{Center for Theoretical Physics, College of Physics, Sichuan University, Chengdu, 610065, China}
\begin{abstract}
    In this work, we study the shadow of Born-Infeld (BI) black holes with magnetic monopoles and Schwarzschild black holes immersed in the BI uniform magnetic field. Illuminated by a celestial sphere, black hole images are obtained by using the backward ray-tracing method. For magnetically charged BI black holes, we find that the shadow radius increases with the increase of nonlinear electromagnetics effects. For Schwarzschild black holes immersed in the BI uniform magnetic field, photons tend to move towards the axis of symmetric, resulting in stretched shadows along the equatorial plane. 
\end{abstract}
\maketitle
\section{Introduction}
The Event Horizon Telescope (EHT) collaboration unveiled the first image of the supermassive black hole in the center of galaxy M87 \cite{Akiyama:2019bqs, Akiyama:2019brx, Akiyama:2019cqa, Akiyama:2019eap, Akiyama:2019fyp, Akiyama:2019sww}. It promotes the current theoretical study of black holes based on modern astronomical observations, and corroborates the direct observations of black hole images. Just recently, the EHT released the image of the supermassive black hole Sgr A* in the center of the Milky Way galaxy \cite{Akiyama:2022aki1, Akiyama:2022aki2, Akiyama:2022aki3, Akiyama:2022aki4, Akiyama:2022aki5, Akiyama:2022aki6}, which confirms that this giant object is consistent with a Kerr black hole and provides valuable clues about it. 

Black hole images provide direct evidence of their existence, revealing much information about black holes and their surroundings. In particular, the dark area inside a black hole image is called the black hole shadow, which can give us information about the basic properties of black holes and serve as a useful tool to test general relativity. Synge first discussed the shadow of Schwarzschild black holes in Ref. \cite{Synge:1966okc}. Bardeen \cite{Bardeen:1972fi} soon studied the shadow cast by Kerr black holes. In recent years, this topic has been extended to other black holes by various researchers \cite{Cunha:2016bpi, Cunha:2015yba, Hioki:2009na, Takahashi:2005hy, Gan:2021xdl, Tsukamoto:2017fxq, Wei:2013kza, Abdujabbarov:2012bn, Schee:2008kz, Podolsky:2012he, Bambi:2011yz, Bambi:2010hf, Younsi:2016azx, Cunha:2016wzk, Belhaj:2021rae, Saha:2018zas, Eiroa:2017uuq, Amir:2016cen, Abdujabbarov:2016hnw, Lara:2021zth, He:2021aeo, Cunha:2018acu, Li:2020drn, Bohn:2014xxa, Guo:2022muy, Lee:2021sws, Kim:2019hfp, Bambi:2019tjh, Vagnozzi:2019apd, Vagnozzi:2020quf, Khodadi:2020jij, Vagnozzi:2022moj}.

After reporting the first black hole image , the EHT cooperations have conducted in-depth studies of previously collected data on the M87*, revealing a new view of the black hole in polarized lights \cite{EventHorizonTelescope:2021bee, EventHorizonTelescope:2021srq}. The polarized black hole images indicate a signature of extreme magnetic fields around black holes. These observations provide new information about the magnetic field structure outside the black hole. With the astronomical observations of strong magnetic fields around black holes, influences of magnetic fields on shadow images have been studied \cite{Junior:2021dyw}. Moreover, effects of nonlinear electrodynamics on black hole shadows were investigated \cite{Okyay:2021nnh, Hu:2020usx, Zhong:2021mty, Allahyari:2019jqz}.

For black hole shadows in nonlinear electrodynamics, it showed that photons travel along null geodesics in a so-called effective geometry \cite{Plebansky:1968} rather than the actual spacetime. This means that in a nontrivial vacuum, the motion of lights can be viewed as electromagnetic waves propagating through a classical dispersive medium \cite{Dittrich:1998fy}. The medium causes corrections to the equations of motion described in nonlinear dynamics \cite{Shore:1995fz}. In recent years, with the development of nonlinear quantum electrodynamics, the study of light propagation has also made remarkable progress, e.g., the geometry of light propagation in nonlinear electrodynamics \cite{Cai:2004eh} and the deflection of magnet star in Born-Infeld \cite{Kim:2022xum}. 

To avoid the occurrence of singularities in the Maxwell theory, Born and Infeld proposed the Born-Infeld (BI) nonlinear electrodynamics model \cite{Born:1934gh}, which considered the action principle of free particles in relativistic theories, and naturally imposed an upper bound on the velocity of particles. Surprisingly, the BI electrodynamics can be derived from string theory in a low energy limit \cite{Fradkin:1985qd, Tseytlin:1986ti}. Later, some works demonstrated that BI electrodynamics could govern the dynamics of D branes and derive some soliton solutions of hypergravity \cite{Wiltshire:1988uq, Salazar:1987ap}. In recent years, these ideas were applied to modify the Einstein-Hilbert action in the gravity theory, which has attracted extensive attention \cite{Hanazawa:2019kbi, Olszewski:2015wva, VanProeyen:2013bia, Hoffmann:1935ty, Callan:1997kz, Cataldo:1999wr, Garcia-Salcedo:2000ujn, Fernando:2003tz, Miskovic:2008ck, Tseytlin:1997csa, Cecotti:1986gb, Jing:2020sdf, Wang:2020ohb, Gan:2019jac, Liang:2019dni, Tao:2017fsy, Wang:2019kxp, Bi:2020vcg, Delhom:2019zrb, BeltranJimenez:2021oaq}. 

In this paper, we are interested in how BI electrodynamics will affect black hole shadows. We first study the static spherically symmetric BI black hole with magnetic monopoles and obtain its metric function. The trajectories of photons are derived by introducing the effective metric. Based on the backward ray-tracing method, the black hole images are obtained with different magnetic charges and BI parameter. Moreover, we give the first-order perturbative effective metric for Schwarzschild black holes immersed in a BI uniform magnetic field and study shadows with various parameters of the BI nonlinear magnetic effects. Unique characteristics have been found, which may be verified by future observations for black holes with strong magnetic fields.

The organization of this paper is as follows: In section II, we derive the effective metric for BI black holes with magnetic monopoles and investigate the images of their shadows. In section III, we obtain the perturbative effective metric for black holes immersed in the BI uniform magnetic field, and analyze the influence of the nonlinear magnetic field on shadows. Finally, we summarize our results in section IV. In addition, we introduce the numerical backward ray-tracing method in Appendix \ref{ray-tracing Method}. In this paper, we set $16\pi G=c=1$ for simplicity.

\section{BI Black holes with magnetic monopoles}
In this section, we first obtain the static and spherically symmetric solution for BI black holes with magnetic monopoles. Consider an Einstein-Born-Infeld action \cite{Cai:2004eh}
\begin{align}
    S=\int d^4x\sqrt{-g}\left[\mathcal{R}+L\left(F\right)\right],
    \label{action}
\end{align}
where $\mathcal{R}$ is the scalar curvature, $F=F ^{\mu \nu}F_{\mu \nu}$, and
\begin{align}
    L\left(F\right)=4 \beta^{2}\left(1-\sqrt{1+\frac{F}{2 \beta^2}}\right)
    \label{L}
\end{align}
is the BI Lagrangian. The parameter $\beta$ is called the BI parameter with the dimension of mass. In the limit $\beta\rightarrow \infty$, $L(F)$ reduces to the standard Maxwell form. The Einstein-Born-Infeld equations can be obtained by varying the action with respect to the gauge field $A_\mu$ and the metric $g_{\mu\nu}$, yielding
\begin{align}
    \partial_{\mu}\left(\frac{\sqrt{-g} F ^{\mu \nu}}{\sqrt{1+\frac{F}{2 \beta^{2}}}}\right)=0,
    \label{motion of ele}
\end{align}
\begin{align}
    \mathcal{R}_{\mu \nu}-\frac{1}{2} \mathcal{R} g_{\mu \nu}=\frac{1}{2} g_{\mu \nu} L(F)+\frac{2 F_{\mu \alpha} F_{\nu}^{\alpha}}{\sqrt{1+\frac{F}{2 \beta^{2}}}}.
    \label{motion of Ein}
\end{align}
\begin{figure}[t]
    \centering
    \includegraphics[width=8cm]{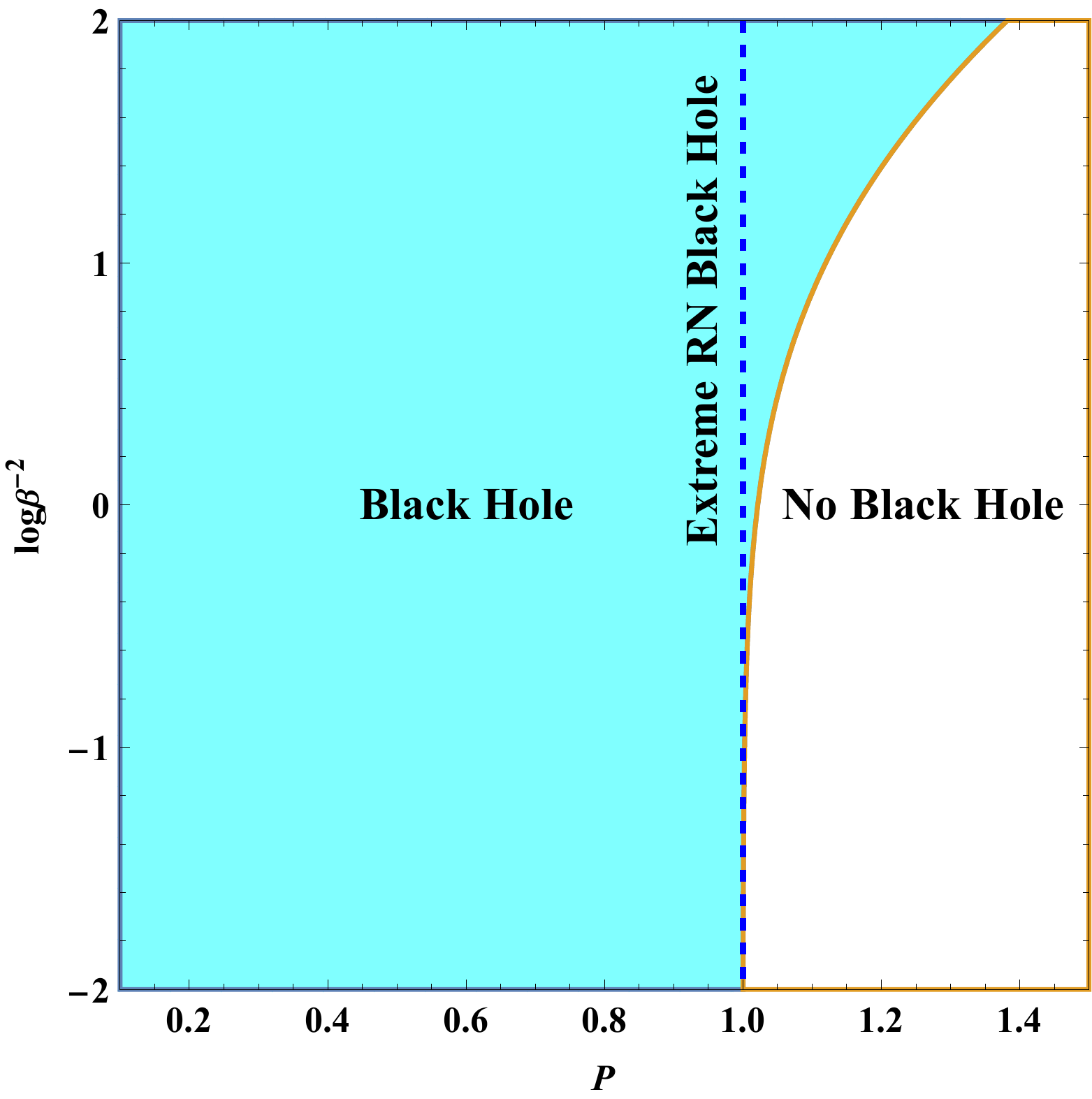}
    \caption{Region plot in the $P-\log\beta^{-2}$ parameter space for magnetically charged BI black holes with $m=1$. Black holes exist in the cyan region, and the blue dashed line stands for the extreme RN black hole. With the decrease of $\beta$, the upper limit of the magnetic charge $P$ increases.}
    \label{parameters space of mo}
\end{figure}

We consider the static and spherically symmetric ansatz taking the form as
\begin{align}
    ds^{2}=-f(r) d t^{2}+\frac{1}{f(r)} d r^{2}+r^{2} (d \theta^{2}+ \sin ^{2} {\theta} d \phi^{2}),
    \label{metric}
\end{align}
\begin{align}
    F_{\theta\phi}=-F_{\phi\theta}=P\sin{\theta},
    \label{MUansatz}
\end{align}
where $P$ is a positive constant related to the magnetic charge. This gauge field is generated by magnetic monopoles, which satisfies Eq. (\ref{motion of ele}). By substituting Eqs. (\ref{metric}) and (\ref{MUansatz}) into Eq. (\ref{motion of Ein}), the field equation can be rewritten as 
\begin{align}
    rf^{\prime}(r)+\frac{r^{2}}{2} f^{\prime \prime}(r)=2 r^{2} \beta^{2}\left(1-\sqrt{1+\frac{P^{2}}{\beta^{2} r^{4}}}\right)+\frac{2 P^{2}}{r^{2} \sqrt{1+\frac{P^{2}}{\beta^{2} r^{4}}}},
    \label{MMeq}
\end{align}
where a prime stands for the derivative with respect to the coordinate $r$. From Eq. (\ref{MMeq}), we obtain the metric function for magnetically charged BI black holes
\begin{align}
    f(r)= 1-\frac{2m}{r}+\frac{2\beta ^2 r^2}{3}-\frac{2}{3} \beta ^2 r^2 \,_2F_1\left(-\frac{3}{4},-\frac{1}{2};\frac{1}{4};-\frac{P^2}{r^4 \beta ^2}\right),
    \label{MMmf}
\end{align}
where $m$ stands for the black hole mass, and $_2F_1$ is the hypergeometric function. This solution can be regarded as the dual form for the electrically charged BI black hole metric obtained in Refs. \cite{Cai:2004eh, Cataldo:1999wr}. In the limit $\beta\rightarrow \infty$, the metric function in Eq. (\ref{MMmf}) reduces to the one for the Reissner-Nordström (RN) black hole. In Fig. \ref{parameters space of mo}, we plot the $\log$ parameter space about $P$ and $\beta$. When the BI nonlinear effect is weak, it almost coincides with the RN black hole. With the decrease of $\beta$, the upper limit of magnetic charge $P$ increases.

Due to the nonlinear electrodynamics effects, photons propagate along null geodesics in the effective metric rather than the background metric. The effective metric $G^{\mu\nu}$ takes the form as \cite{Novello:1999pg}
\begin{align}
    G^{\mu\nu}=4L_{FF}F^{\mu}_{\alpha}F^{\alpha\nu}-L_{F}g^{\mu\nu},
    \label{eff metric}
\end{align}
where $L_F=d L/d F$ and $L_{FF}=d^2 L/d F^2$. Substituting Eqs. (\ref{L}), (\ref{metric}) and (\ref{MUansatz}) into Eq. (\ref{eff metric}), we derive the effective metric for the static and spherically symmetric BI black hole with magnetic monopoles
\begin{align}
    ds^{2}_{eff}=\left(1+\frac{P^2}{\beta ^2 r^4}\right)^{\frac{1}{2}}\left[-f(r) d t^{2}+\frac{1}{f(r)} d r^{2}+h(r) (d \theta^{2}+ \sin ^{2} {\theta} d \phi^{2})\right],
    \label{MMeff}
\end{align}
where
\begin{align}
    h(r)=r^2\left(1+\frac{P^2}{\beta ^2 r^4}\right).
\end{align}

The effective geodesic equations are given by a group of eight first-order Hamilton equations. Setting $\lambda$ to be the affine parameter, one can derive the effective geodesic equations as
\begin{align}
    p^{\mu}=\frac{dx^{\mu}}{d\lambda},\quad \frac{dp^{\mu}}{d\lambda}+\Gamma^{\mu}_{\rho \sigma}p^{\rho}p^{\sigma}=0,
    \label{geodesic equation}
\end{align}
where $p^{\mu}$ is the 4-momentum vector of a photon, and $\Gamma^{\mu}_{\rho \sigma}$ is the affine connection of the effective metric. It is convenient to define a new dual vector $q_{\mu}\equiv G_{\mu \nu}p^{\nu}$. Photons have two Killing vectors corresponding to the conserved energy $E=-q_t$ and angular momentum $L=q_{\phi}$ \cite{Hu:2020usx,Zhong:2021mty}. By substituting the effective metric (\ref{MMeff}) into the Hamiltonian constraint $G^{\mu\nu}q_{\mu}q_{\nu}=0$, we obtain
\begin{align}
    f(r)q_r^2-\frac{1}{f(r)}E^2+\frac{1}{h(r)}\left(q_{\theta}^2+L^2\right)=0,
\end{align}
where we can introduce the effective potential $V$ of photons
\begin{align}
    V(r)=\frac{L^2}{h(r)}-\frac{E^2}{f(r)}.
    \label{V}
\end{align}
For the critical case $V(r_{\text{ph}})=V'(r_{\text{ph}})=0$, photons can orbit around the spherically symmetric black hole at a constant radius $r_{\text{ph}}$, which forms a sphere called the photon sphere. Being an intrinsic property of the black hole and independent of the observer, it determines the boundary for photons to fall into a black hole or escape to infinity.

For a distant observer located at $r_{o}\gg m$, the shadow radius is given by $r_{\text{sh}}=r_{o}\sin\theta\simeq L/E|_{r_{\text{ph}}}$. The parameter $\theta$ represents the inclination angle of the photons emitted from $r=r_{\text{ph}}$. By substituting the critical condition into Eq. (\ref{V}), the shadow radius takes the form as
\begin{align}
    r_{\text{sh}}=\pm\sqrt{\frac{h(r_{\text{ph}})}{f(r_{\text{ph}})}}.
    \label{rph}
\end{align}
\begin{figure}[t]
    \begin{center} 
        \subfigure[\;$\beta=1$]{\includegraphics[width=5.25cm]{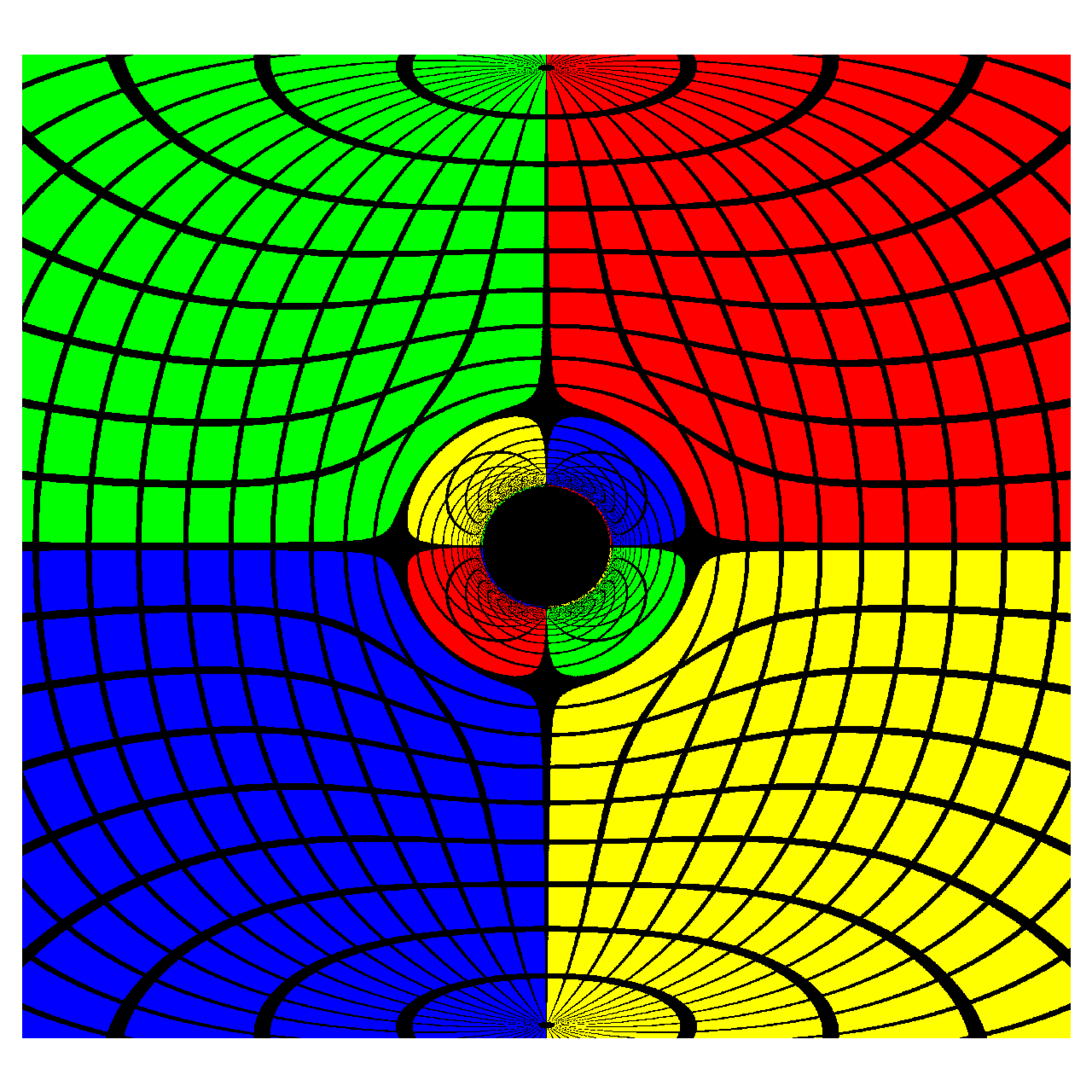}
        \label{monopoleShadow11}}
        \subfigure[\;$\beta=0.1$]{\includegraphics[width=5.25cm]{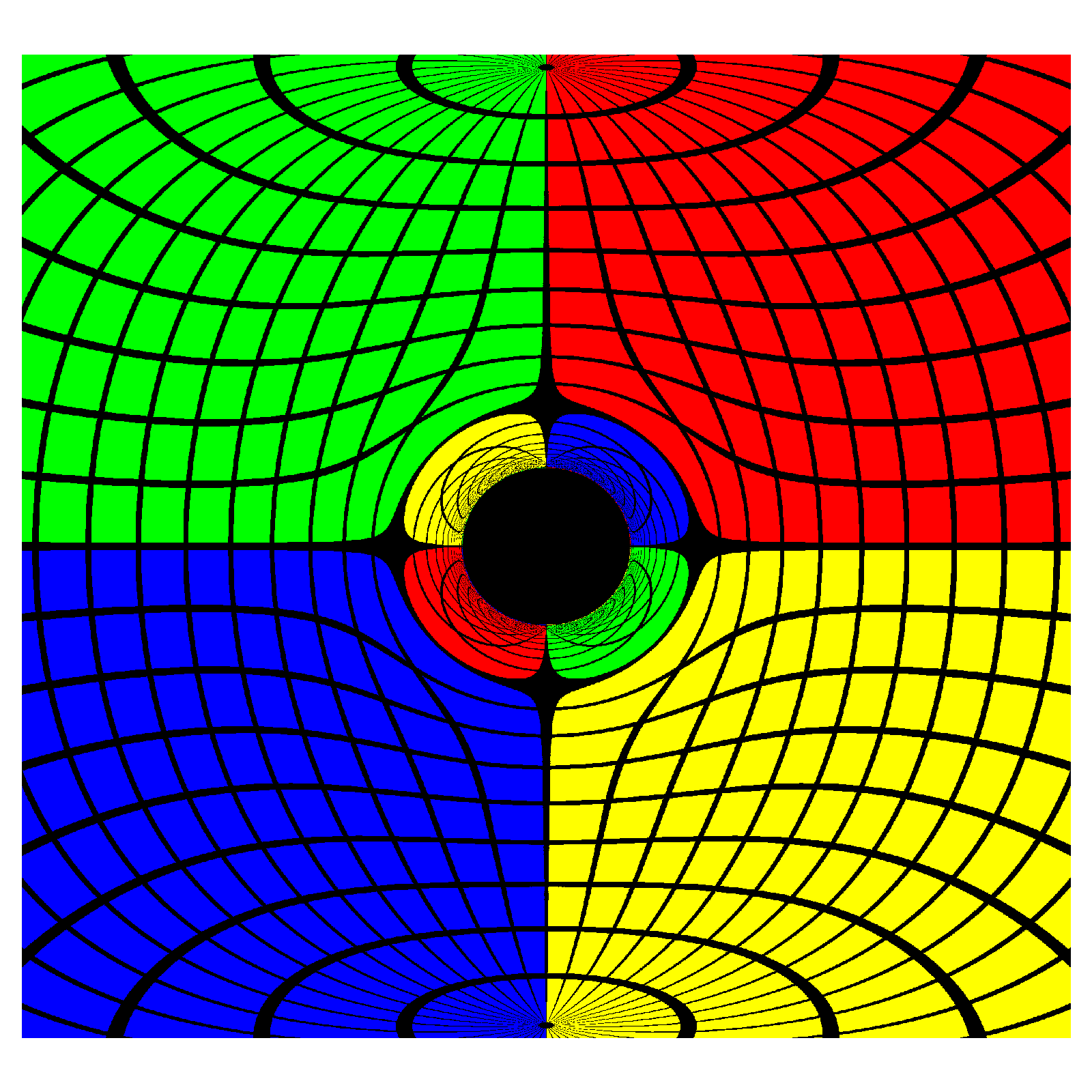}
        \label{monopoleShadow12}}
        \subfigure[\;$\beta=0.01$]{\includegraphics[width=5.25cm]{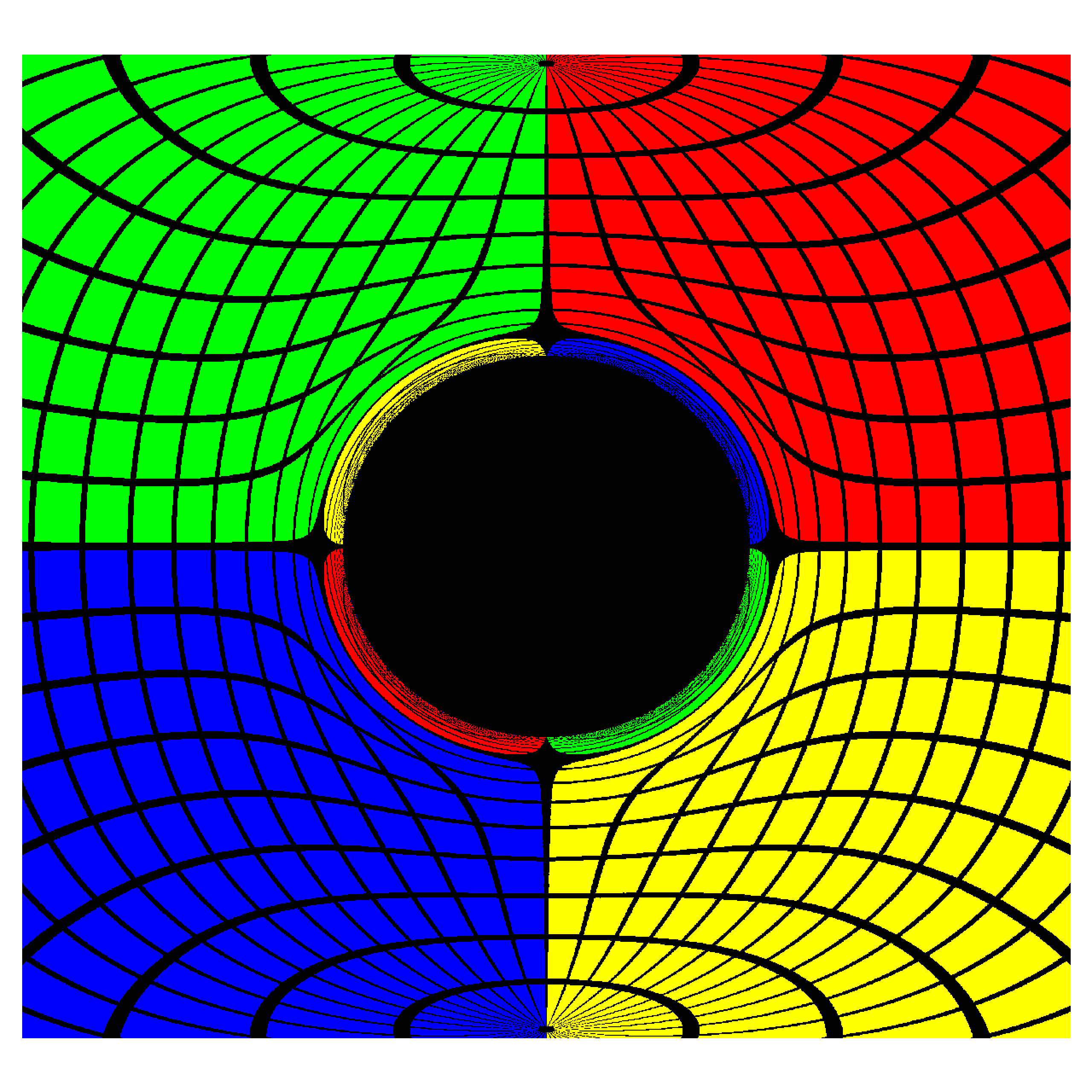}
        \label{monopoleShadow13}}
        \caption{Images of magnetically charged BI black holes with $m=1$ and $P=0.8$ for $\beta=1$, $\beta=0.1$ and $\beta=0.01$ respectively. We find that the shadow radius increases with the decrease of $\beta$.}
        \label{monopoleShadow1}
    \end{center}
\end{figure}

Based on the effective geodesic equations in Eq. (\ref{geodesic equation}) and using the numerical backward ray-tracing method  programmed with Wolfram$\circledR$Mathematica, we plot $2000\times 2000$ pixels images for magnetically charged BI black holes with $m=1$ and $P=0.8$ for $\beta=1$, $\beta=0.1$ and $\beta=0.01$ in Fig. \ref{monopoleShadow1}, respectively. It shows that the radius of the shadow increases significantly when $\beta$ decreases to the order of $10^{-2}$, and the surrounding gravitational lensed image also becomes thinner with the decrease of $\beta$.
\begin{figure}[htbp]
    \begin{center}
        \subfigure[\;$r_{\text{sh}}-\beta$]{\includegraphics[width=7.5cm]{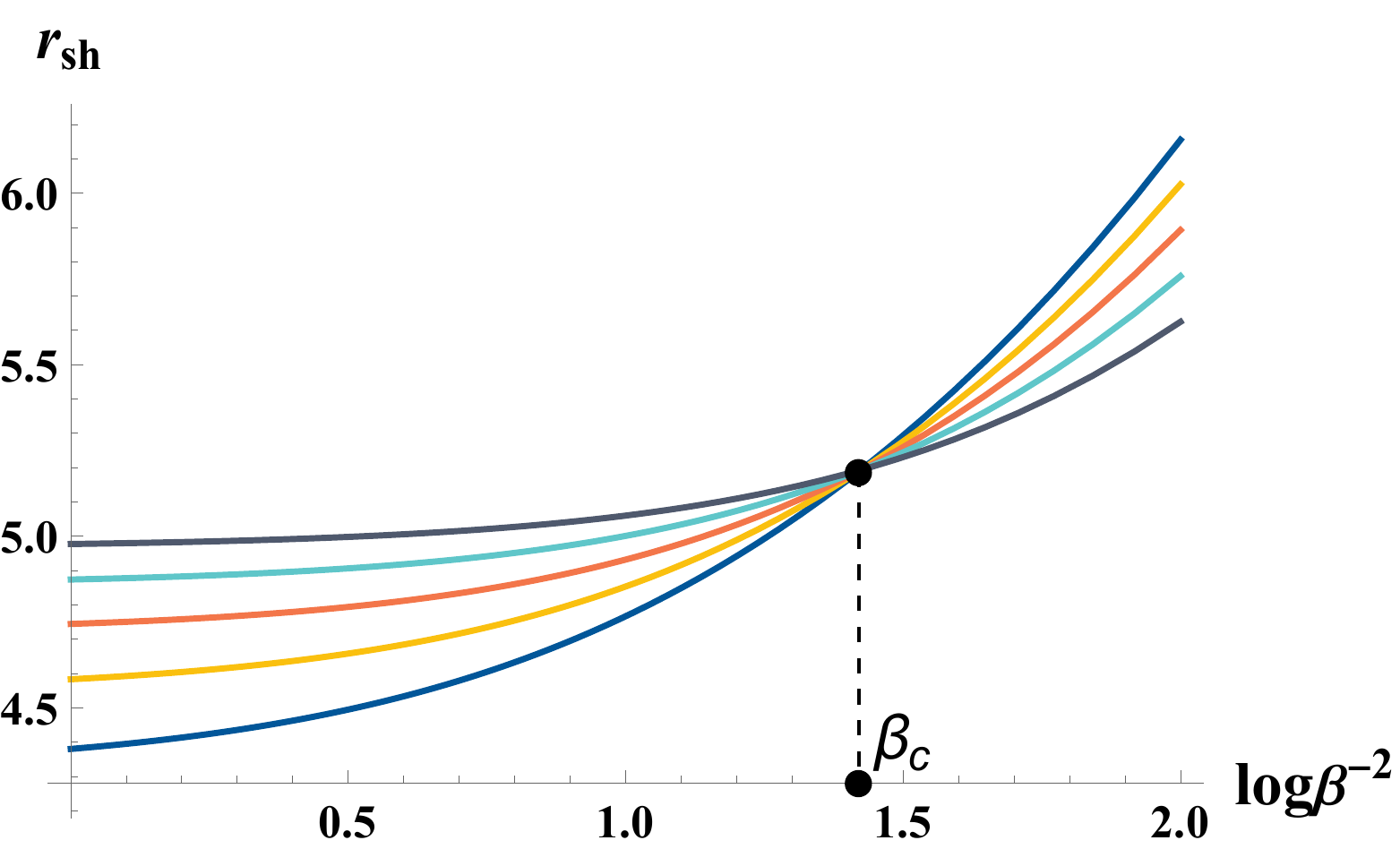}
        \label{bc1}}
        \subfigure[\;$r_{\text{ph}}/r_h-\beta$]{\includegraphics[width=7.5cm]{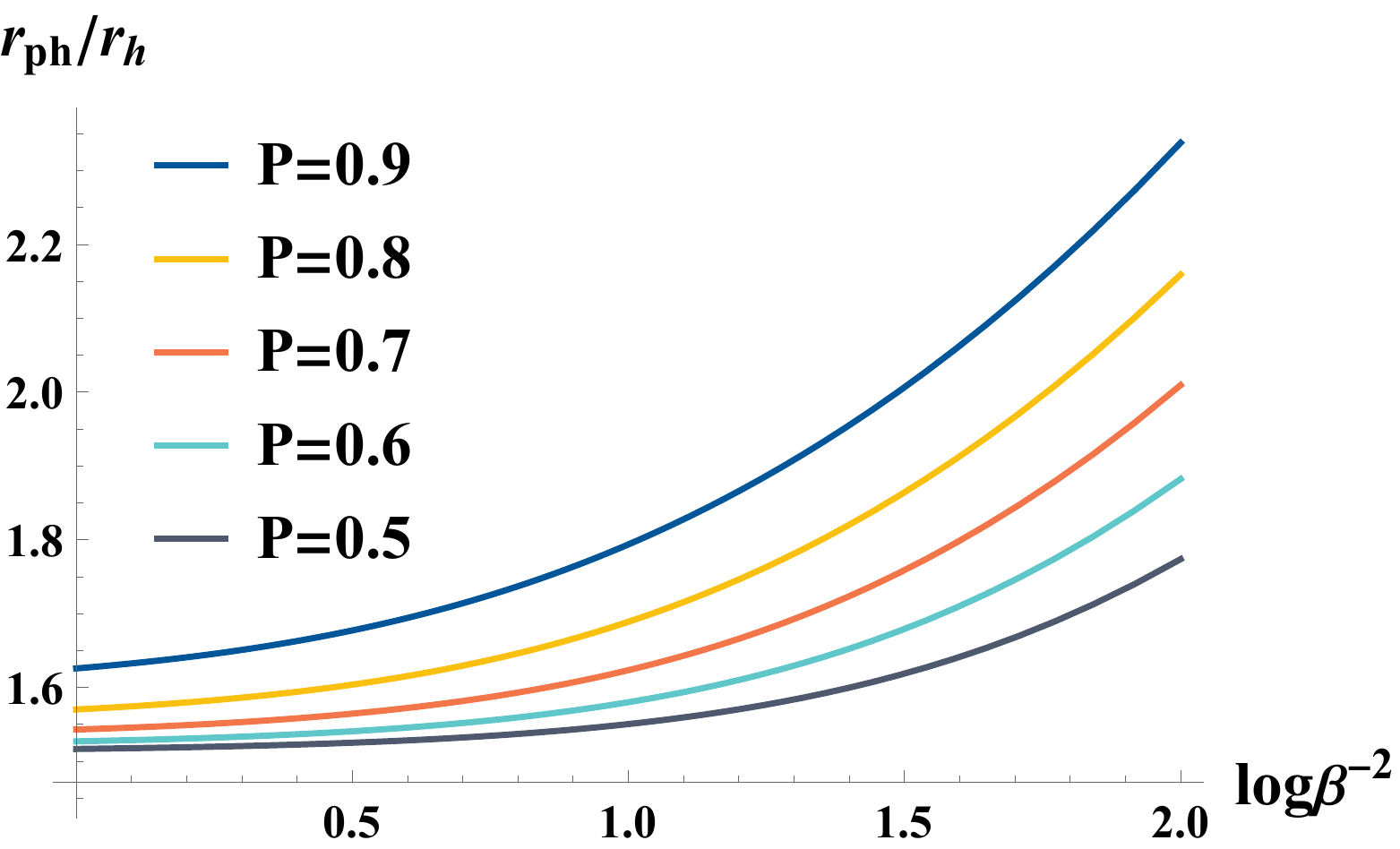}
        \label{rphrh}}
        \caption{The shadow radius $r_{\text{sh}}$ in (a) and the photon sphere radius divided by the horizon radius $r_{\text{ph}}/r_h$ in (b) with respect to the BI parameter $\beta$ for magnetically charged BI black holes for different $P$ with $m=1$. When $\beta=\beta_c$, the radius of the shadow remains unchanged regardless of the value of $P$. While the decrease of $\beta$ or the increase of $P$ makes $r_{\text{ph}}/r_h$ increase monotonically.}
        \label{bc}
    \end{center}
\end{figure}
In addition, the numerical results of the shadow radius $r_{\text{sh}}$ and the photon sphere radius divided by the horizon radius $r_{\text{ph}}/r_h$ are displayed with various BI parameter $\beta$ under different magnetic charge $P$ in Fig. \ref{bc}. In Fig. \ref{bc1}, the shadow radius increases monotonically with $\beta$ decreases. It is worth noting that there is a special $\beta_c\simeq 0.194798$, at which $r_{\text{sh}}$ remains the same regardless of the value of $P$. When $\beta>\beta_c$ ($\beta<\beta_c$), black holes with larger magnetic charge have smaller (larger) shadows. At the same time, the parameter $r_{\text{ph}}/r_{h}$ increases monotonically as $P$ increases or $\beta$ decreases.

\section{Schwarzschild Black holes immersed in BI uniform magnetic fields}
First, we study the effective metric for Schwarzschild black holes immersed in the BI uniform magnetic field. Consider a gauge field for a uniform magnetic field,
\begin{align}
    A_\mu d x^\mu = A_\phi(r,\theta)d\phi,
    \label{ansatz}
\end{align}
where we take the gauge choice $A_r = 0$, and $A_t = 0$ for the absence of the electric field. We assume that the gauge field is axisymmetric, and hence $A_{\phi}$ is independent of $\phi$. Substituting Eq. (\ref{ansatz}) into Eq. (\ref{motion of ele}), we obtain
\begin{align}
    \partial_r\left[\frac{(r-2m)\partial_r A_{\phi}(r,\theta)}{r\sin\theta \sqrt{1+\frac{r(r-2m) \partial_r^2A_{\phi}(r,\theta)+\partial_{\theta}^2A_{\phi}(r,\theta)}{\beta^2r^4 \sin^2{\theta}}}}\right]+\partial_{\theta}\left[\frac{\partial_{\theta} A_{\phi}(r,\theta)}{r^2\sin\theta \sqrt{1+\frac{r(r-2m) \partial_r^2A_{\phi}(r,\theta)+\partial_{\theta}^2A_{\phi}(r,\theta)}{\beta^2r^4 \sin^2{\theta}}}}\right]=0.
    \label{PDE}
\end{align}
In the limit $\beta\rightarrow \infty$, an asymptotic solution could be derived
\begin{align}
    \lim_{\beta \to \infty}  A_{\phi}(r,\theta)\rightarrow\frac{B}{2}r^2\sin^2 {\theta},
    \label{ansatzsol of min}
\end{align}
which describes an axisymmetric magnetic field with strength $B$ along z-axis. This asymptotic gauge field is the analytical solution of the Maxwell uniform magnetic field in Schwarzschild black holes \cite{Wald:1974umf}. On the other hand, since Eq. (\ref{PDE}) is a nonlinear partial differential equation, obtaining the analytical solution of $A_{\phi}(r,\theta)$ is challenging. In this paper, we use perturbative results from Ref. \cite{Bokulic:2019kcc} to give the effective metric for photons moving around Schwarzschild black holes immersed in BI uniform magnetic fields. In fact, the first-order perturbative gauge field takes the form
\begin{align}
    A_{\phi}(r,\theta)=\frac{B}{2}r^2\sin^2 {\theta}+\frac{mB^3}{16\beta^2}\left[4(2r-5m)\cos{2\theta}-(2r-m)(3+\cos{4\theta})\right].
    \label{perturbationansatz}
\end{align}
Substituting Eq. (\ref{perturbationansatz}) into Eq. (\ref{eff metric}) gives the non-zero components of the first-order perturbative effective metric
\begin{figure}[t]
    \begin{center}
        \subfigure[\;$H=0.1$]{\includegraphics[width=5.5cm]{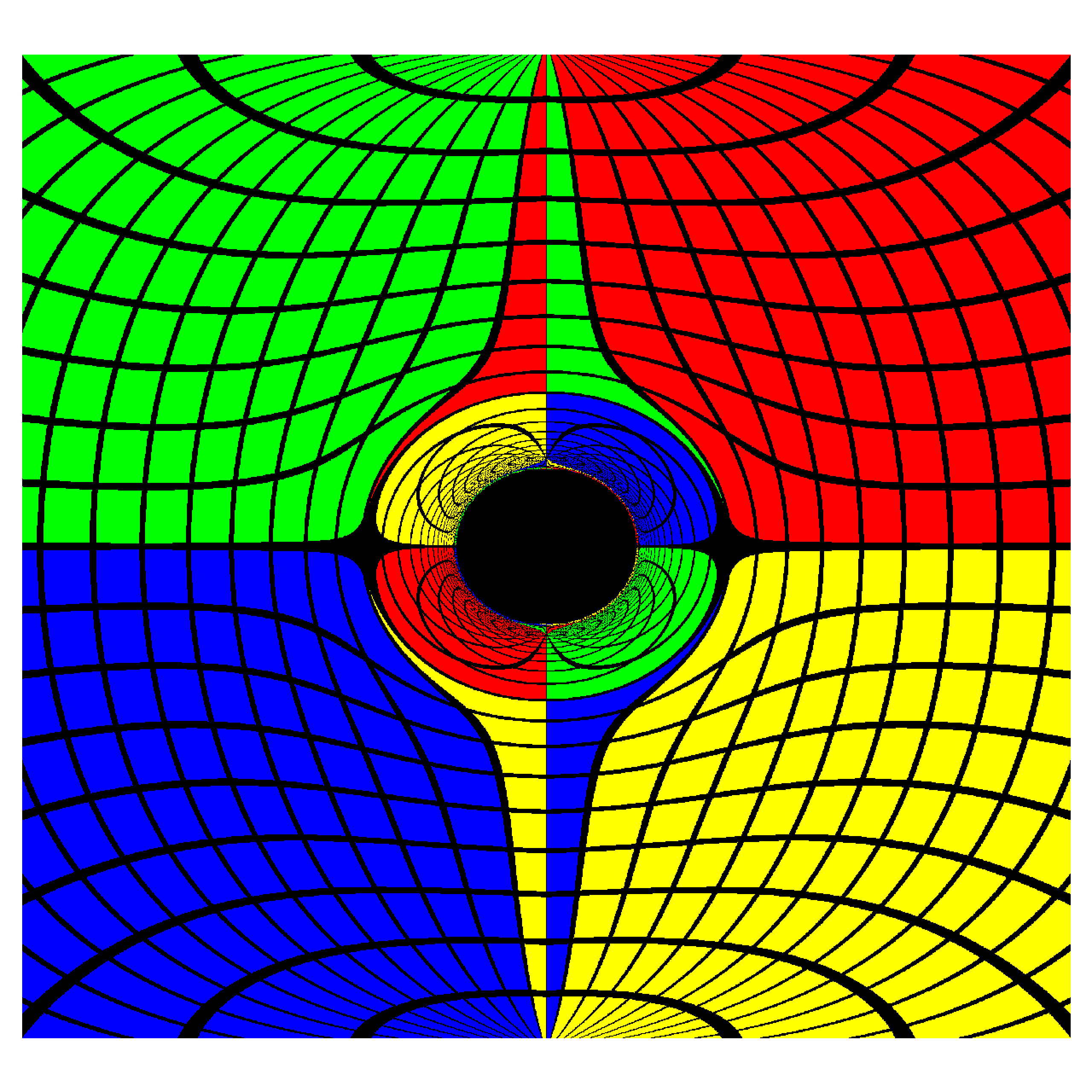}
        \label{UM111}}
        \subfigure[\;$H=0.2$]{\includegraphics[width=5.5cm]{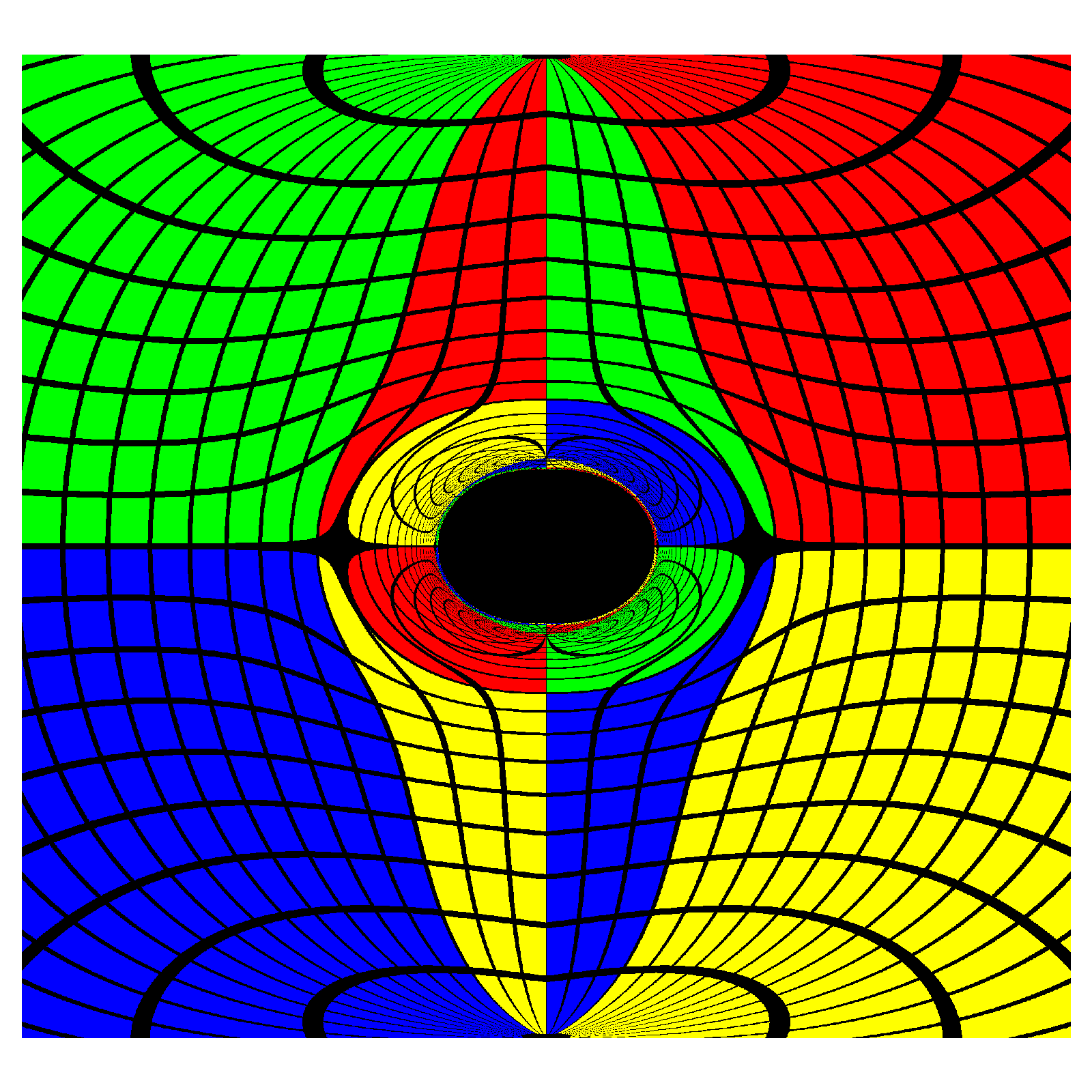}
        \label{UM112}}\\
        \subfigure[\;$H=0.3$]{\includegraphics[width=5.5cm]{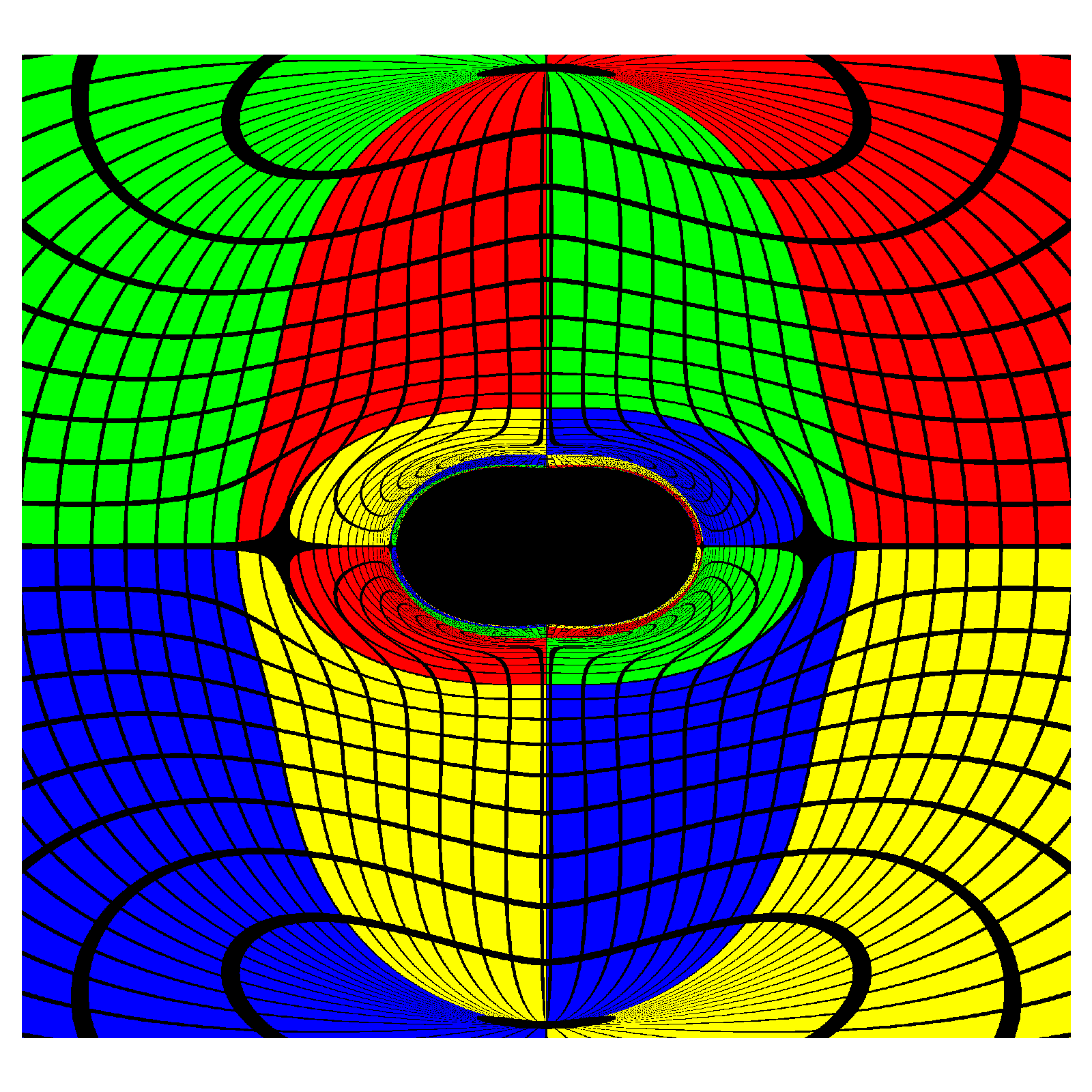}
        \label{UM113}}
        \subfigure[\;$H=0.4$]{\includegraphics[width=5.5cm]{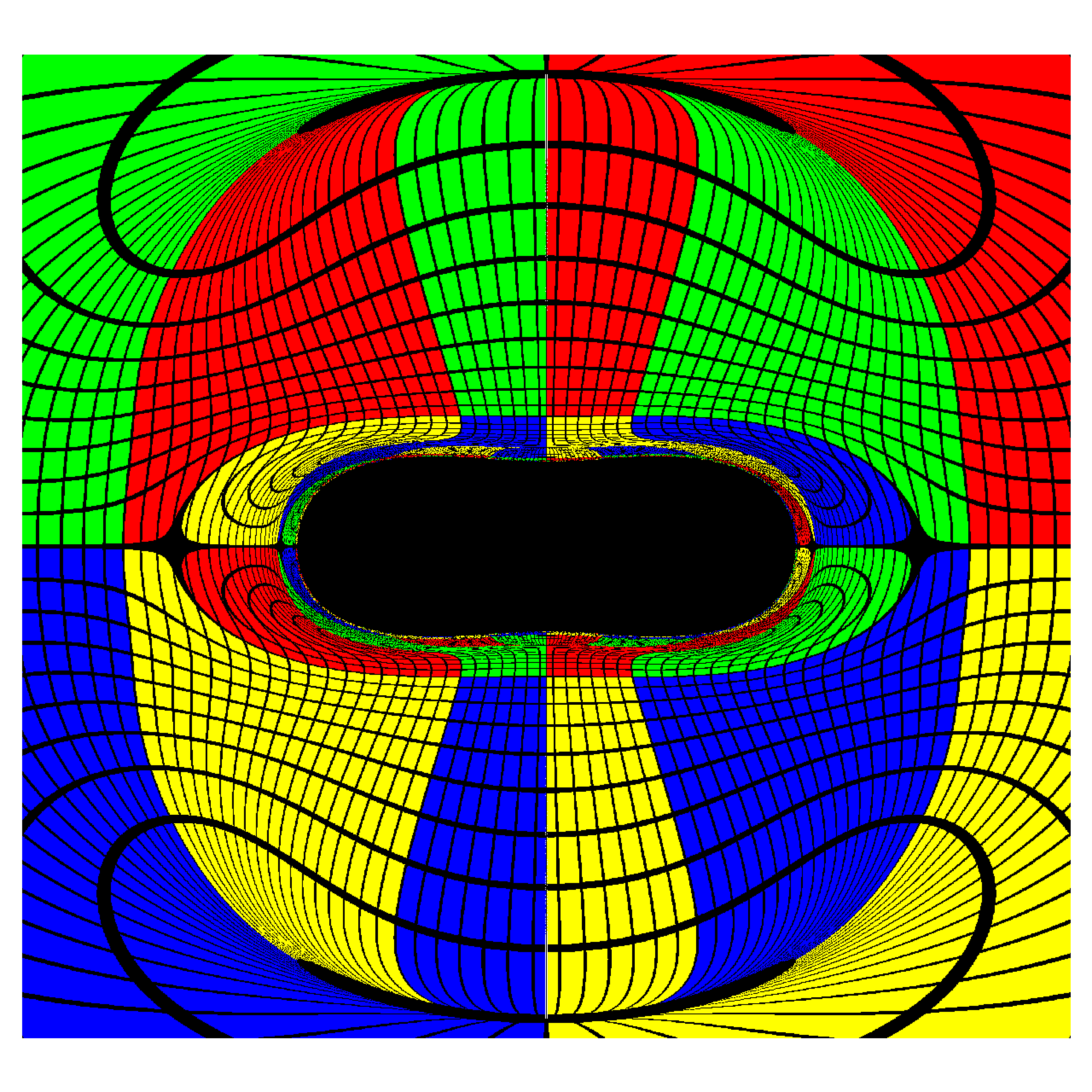}
        \label{UM114}}
        \caption{Images of the Schwarzschild black hole immersed in uniform magnetic fields for different $H$ with the black hole mass $m=1$. With $H$ increases, the shadow contour stretches along the equatorial plane.}
        \label{UM11}
    \end{center}
\end{figure}
\begin{equation}
    \begin{aligned}
        G_{tt}            & =-G_{rr}^{-1}=-\left(1-\frac{2 m}{r}\right)\left[1+\frac{H(r-m+m \cos 2 \theta )}{2 r}\right], \\
        G_{\theta \theta} & =r^2 +\frac{1}{2} H r \left[2 r-m+(m+r)\cos 2 \theta  \right],                            \\
        G_{r\theta}       & =H r \sin \theta  \cos \theta,                                                 \\
        G_{\phi\phi}      & =r^2 \sin ^2\theta +\frac{3}{2} H r \sin ^2\theta  (r-m+m \cos 2 \theta ),
    \end{aligned}
    \label{UMmetric}
\end{equation}
where the dimensionless parameter $H=B^2/\beta^2$ characterizes the strength of the BI effect. 

In Fig. \ref{UM11}, we plot images for the Schwarzschild black hole immersed in the BI uniform magnetic field with $m=1$ for different $H$. Figs. \ref{UM111} and $\ref{UM112}$ display images with $H=0.1$ and $H=0.2$, respectively. Their shadow contours both have tiny stretches along the equatorial plane, whose shapes are similar to ellipses. While for $H=0.3$ in Fig. \ref{UM113} and $H=0.4$ in Fig. \ref{UM114}, shadow contours get more severe stretches, showing shapes similar to peanuts. 
\begin{figure}[t]
    \centering
    \subfigure[]{\includegraphics[height=5cm]{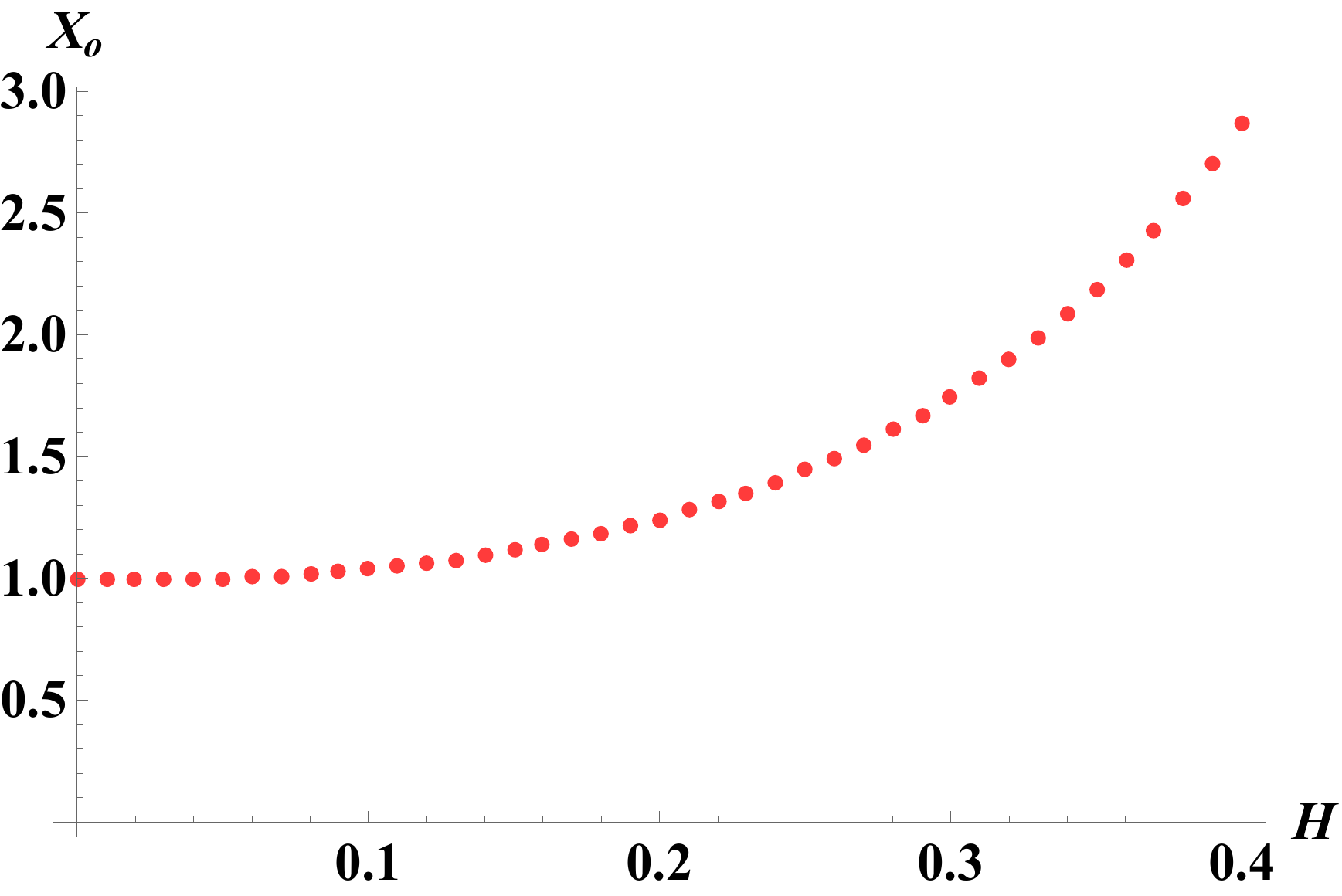}
    \label{XO}}
    \subfigure[]{\includegraphics[height=5cm]{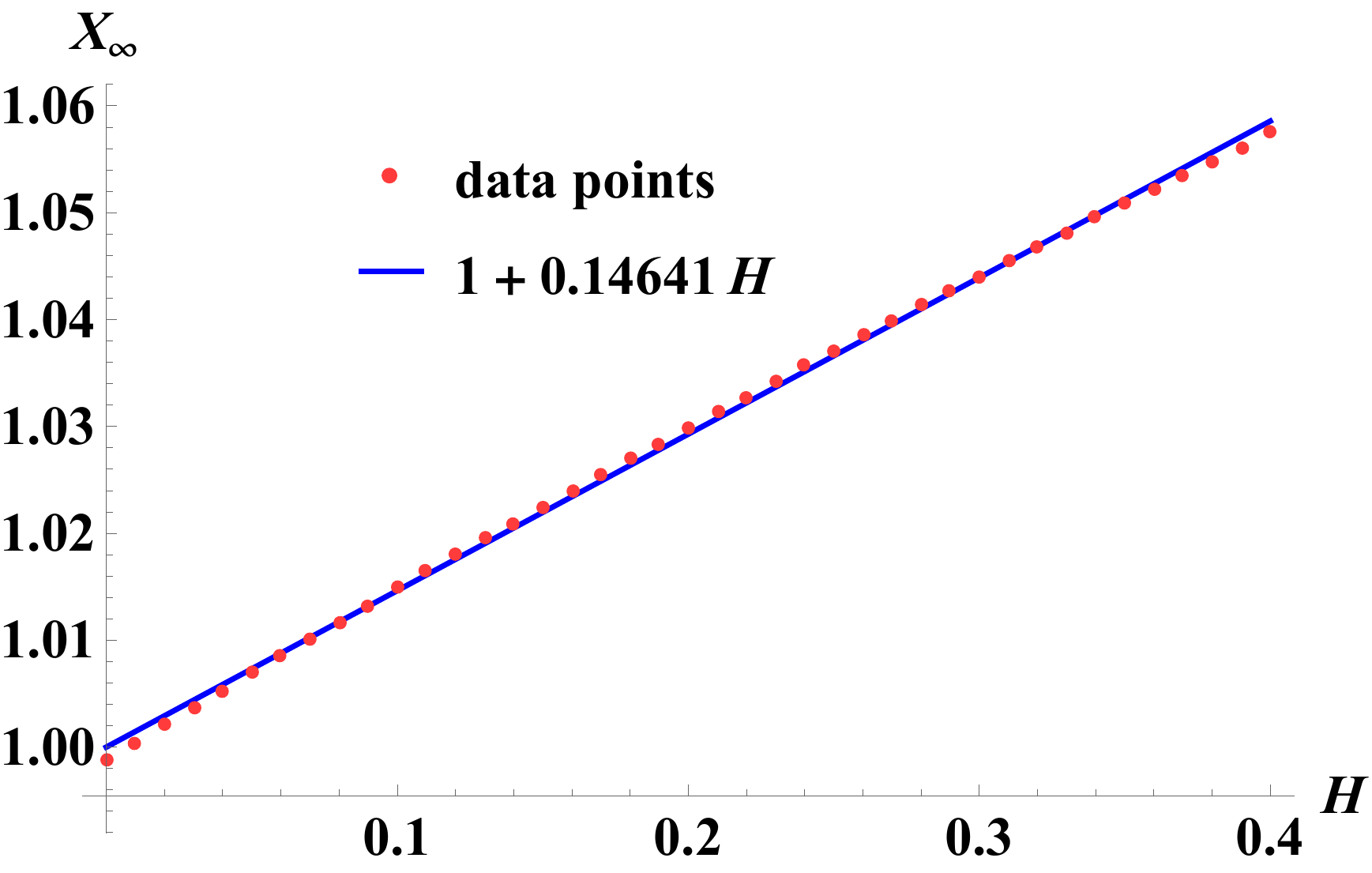}
    \label{XH}}
    \caption{The shadow length along the equatorial plane $X_o$ for an observer located at finite distance in (a), and $X_{\infty}$ for a distant observer in (b). The parameter $X_o$ shows a nonlinear relation with respect to $H$. While $H-X_{\infty}$ fits nicely with a line.}
    \label{X}
\end{figure}

To study how BI nonlinear uniform magnetic field stretches the shadow in detail, we depict the shadow length along the equatorial plane $X_{o}$ for an observer located at finite distance in Fig. \ref{XO}, and $X_{\infty}$ for a distant observer in Fig. \ref{XH}. The values of both $X_o$ and $X_{\infty}$ are divided by the one for black holes with $H=0$. The shadow length $X_o$ shows a nonlinear relationship with $H$ in Fig. \ref{XO}. While for a distant observer in Fig. \ref{XH}, the shadow length shows clearly a linear relation with respect to $H$. 
\begin{figure}[t]
    \begin{center}
        \subfigure[\;Schwarzschild-Maxwell]{\includegraphics[width=5.5cm]{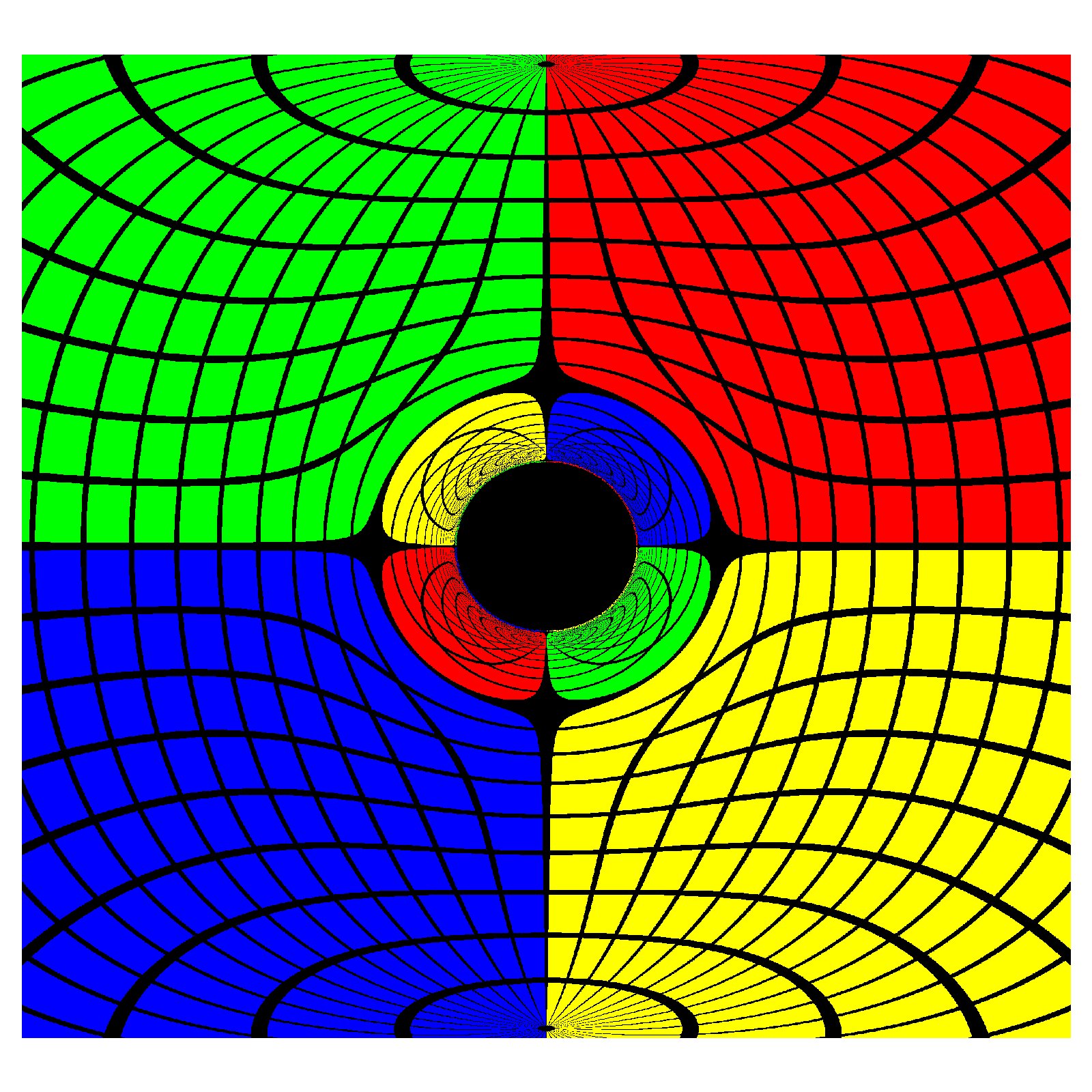}
        \label{UM121}}
        \subfigure[\;Minkowski-Maxwell]{\includegraphics[width=5.5cm]{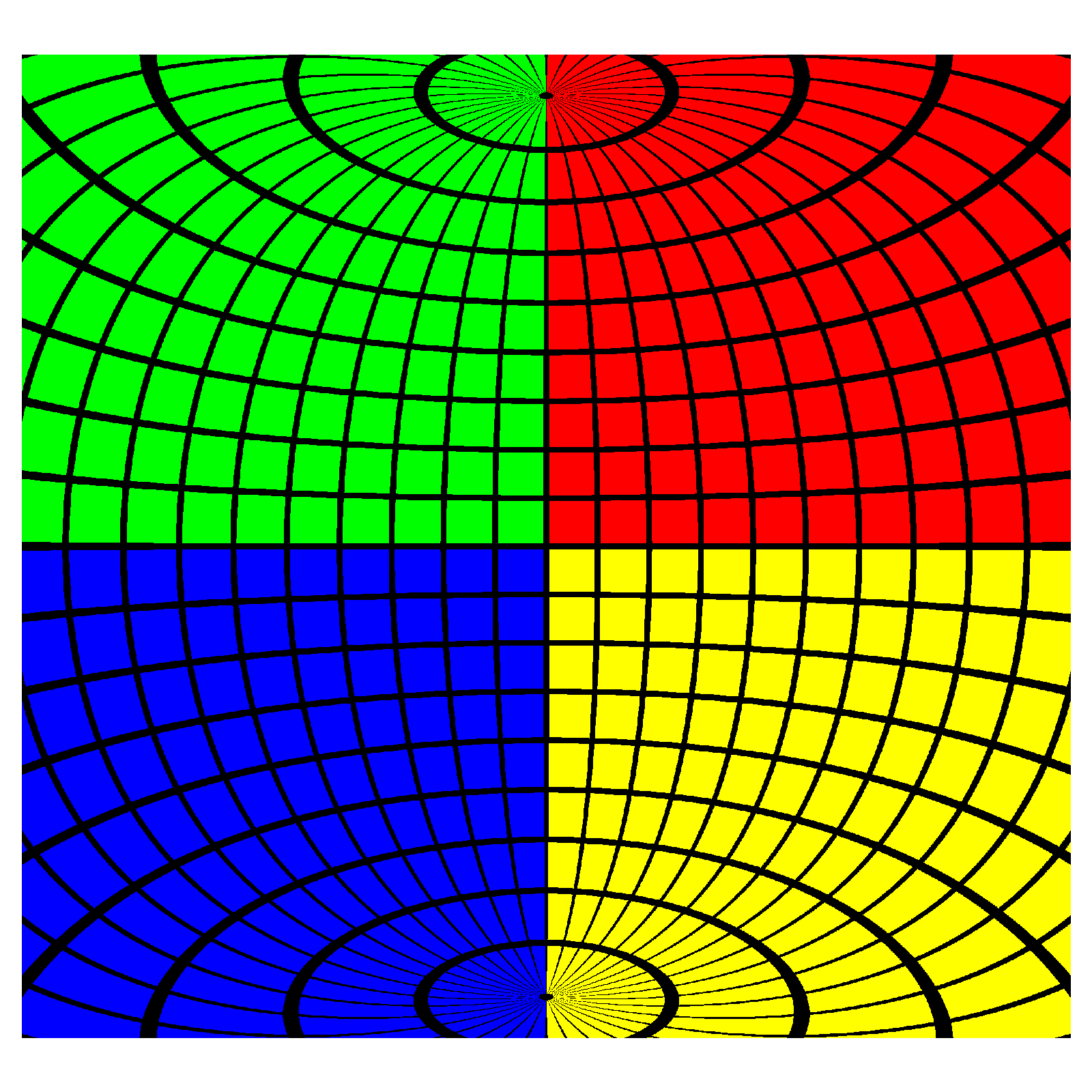}
        \label{UM122}}\\
        \subfigure[\;Schwarzschild-BI]{\includegraphics[width=5.5cm]{BIUnBHH0.4th90n2000f.png}
        \label{UM123}}
        \subfigure[\;Minkowski-BI]{\includegraphics[width=5.5cm]{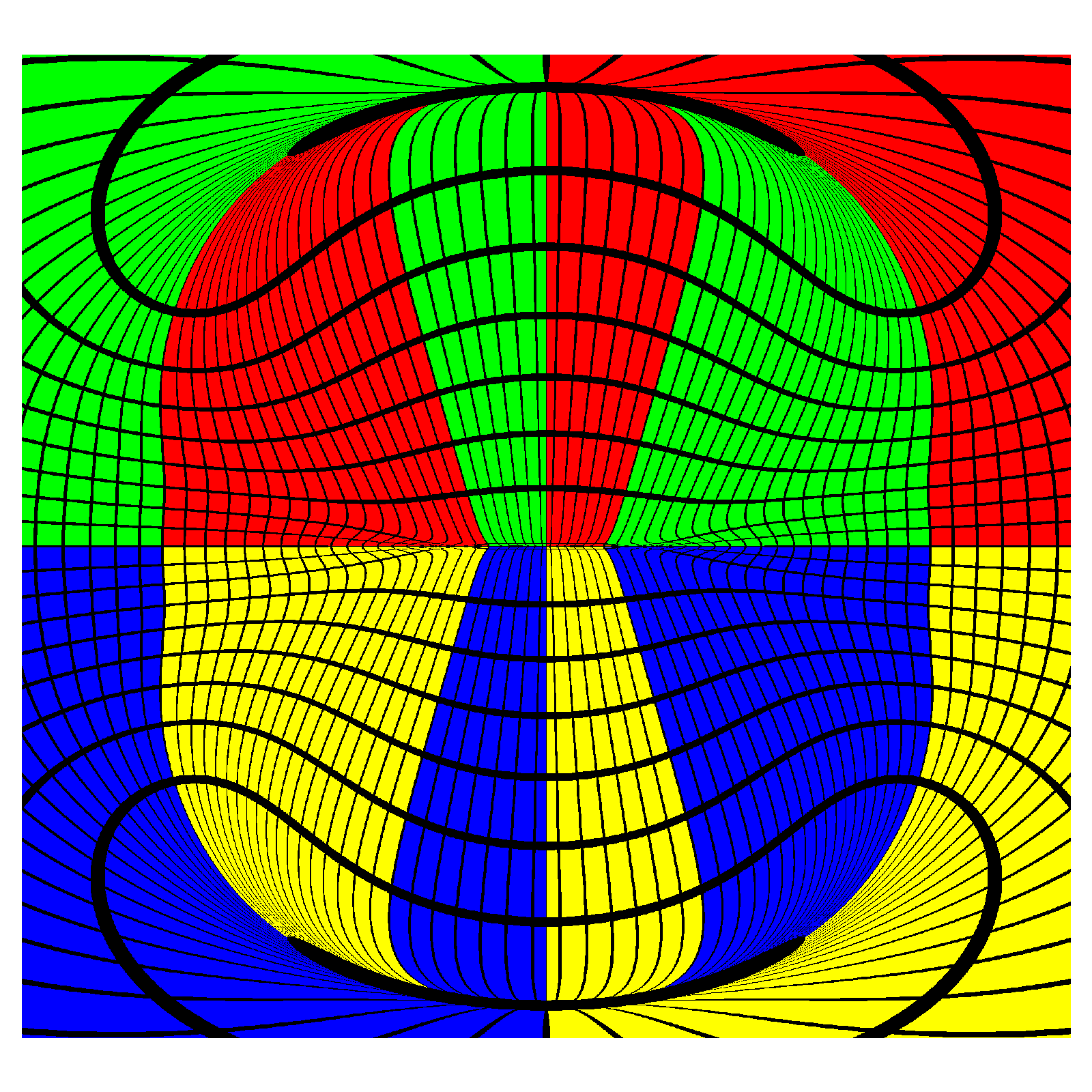}
        \label{UM124}}
        \caption{Images for Schwarzschild black holes (the Minkowski spacetime) in the left panel (right panel), immersed in Maxwell magnetic field (BI magnetic field) on the top panel (bottom panel). We set the black hole mass $m=1$ for (a) and (c), and the BI magnetic field strength $H=0.4$ for (c) and (d). When BI nonlinear effects are considered, axisymmetric higher-orders images of the celestial sphere appear in (c) and (d). It reveals that photons tend to move towards the axis of symmetry under the influence of the nonlinear effects.}
        \label{UM12}
    \end{center}
\end{figure}

To reveal the influence of the BI uniform magnetic field in detail, we further depict image for Schwarzschild black holes immersed in Maxwell magnetic fields in Fig. \ref{UM121}. Without nonlinear effects, Fig. \ref{UM121} shows a circular shadow in the center. While Fig. \ref{UM122} displays a flat Minkowski spacetime immersed in Maxwell magnetic fields. When BI nonlinear effects are considered in Fig. \ref{UM123}, the circular shadow begins to stretch along the equatorial plane. In the meantime, the area outside the shadow gets different image patterns compared with Fig. \ref{UM121}. Fig. \ref{UM124} shows the Minkowski spacetime immersed in the BI magnetic field. By comparing it with Fig. \ref{UM122}, we find that photons no longer travel in a straight line under the influence of nonlinear magnetic field. The image of the red quadrant appears only in the upper right of Fig. \ref{UM122}, while Fig. \ref{UM124} has three red images. Influenced by nonlinear effects, the first-order image of the red quadrant appears on the left side away from the axis, and the second-order image is located on the right side close to the axis. The axisymmetric higher-orders images in Fig. \ref{UM124} reveal that the BI effect makes photons move towards the axis, which can be described by an axial attraction. This axial attraction makes it easier for photons around the equatorial plane to fall into black holes, resulting in a stretched shadow in Fig. \ref{UM123}.

\section{Conclusion}

In this paper, we first studied shadows of BI black holes with magnetic monopoles. By solving the Einstein-Born-Infeld equations, the background metric was obtained. We depicted the parameter space about the magnetic charge $P$ and BI parameter $\beta$ in Fig. \ref{parameters space of mo}. We found that the decrease of $\beta$ increases the upper limit of magnetic charge of the black hole. Based on the numerical backward ray-tracing method, we plotted images for the BI black hole with magnetic monopoles in Fig. \ref{monopoleShadow1}, which shows that the radius of the shadow increases with the decrease of the BI parameter. Moreover, the numerical results of $r_{\text{sh}}$ and $r_{\text{ph}}/r_h$ were displayed in Fig. \ref{bc}. We found a specific $\beta_c$ in which $r_{\text{sh}}$ remains unchanged regardless of the value of $P$. Meanwhile, $r_{\text{ph}}/r_{h}$ increases monotonically as $P$ increases or $\beta$ decreases. 

Next, we investigated shadows of Schwarzschild black holes immersed in the BI uniform magnetic field. By deriving a perturbative solution of the effective metric, we plotted black hole images with different nonlinear magnetic field strengths $H$ in Fig. \ref{UM11}. As $H$ increases, the shadow contour stretches along the equatorial plane. Furthermore, we depicted the shadow length along the equatorial plane $X_{o}$ for an observer located at finite distance in Fig. \ref{XO}, and $X_{\infty}$ for a distant observer in Fig. \ref{XH}. The length $X_o$ shows a nonlinear relationship with $H$. While $X_{\infty}$ as a function of $H$ fits nicely with a line, showing that the increases of nonlinear magnetic fields linearly increase the shadow length along the equatorial plane. We further plotted images for Schwarzschild black holes and the Minkowski spacetime immersed in Maxwell or BI magnetic fields in Fig. \ref{UM12}. By comparing Figs. \ref{UM122} and \ref{UM124}, we found that the photon tends to move towards the axis of symmetry due to the effect of BI uniform magnetic fields. This axial attraction makes it easier for photons around the equatorial plane to fall into black holes, and results in a stretched shadow.

The first-order perturbative solutions derived in this paper reveals some physical characteristics of the BI uniform magnetic field. It also qualitatively explained the deformation of shadows when Schwarzschild black holes is immersed in the nonlinear magnetic field. Though getting the analytical solution of the nonlinear partial differential equation in Eq. (\ref{PDE}) is challenging, it is of great interest to obtain the numerical solution for BI uniform magnetic fields in future works. 

\begin{acknowledgments}
    We are grateful to Guangzhou Guo and Jiayi Wu for useful discussions. This work is supported by NSFC (Grant No.12147207).
\end{acknowledgments}

\appendix
\begin{appendices}
\section{ray-tracing Method}\label{ray-tracing Method}
\begin{figure}[H]
    \begin{center}
        \subfigure[]{
            \includegraphics[height=7cm]{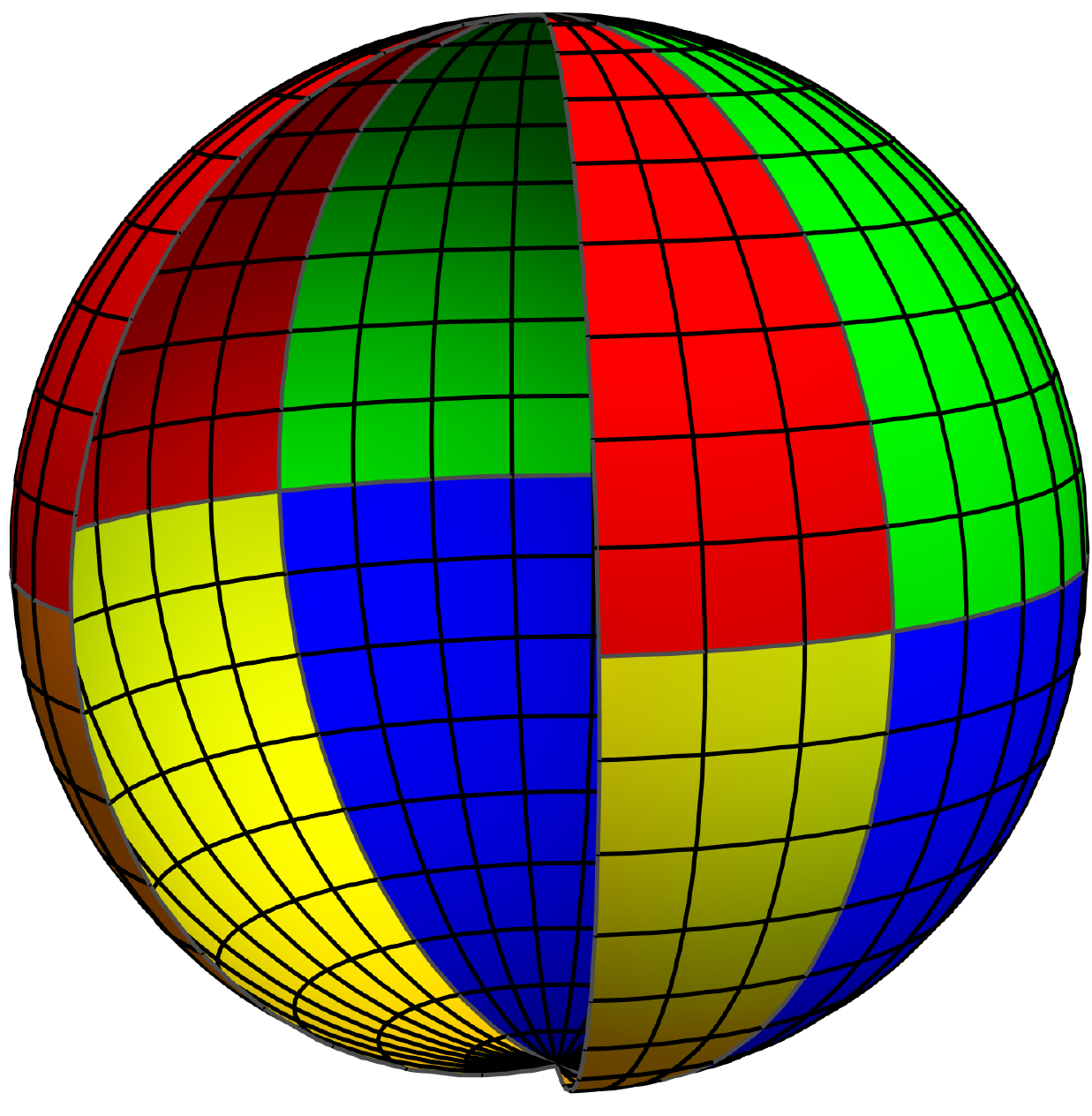}
            \label{3Dcs}}
        \subfigure[]{
            \includegraphics[height=7cm]{MinMaxwell.png}
            \label{cs}}
        \caption{The celestial sphere in (a), and the observational image of the celestial sphere in the Minkowski spacetime in (b).}
    \end{center}
\end{figure}

Celestial sphere is a concept in astrophysics where all objects in the sky can be seen through a projection on the sphere. To compute photon geodesics and get the image of the shadow, we divide the celestial sphere into four parts and give each a different color in Fig. \ref{3Dcs} \cite{Bohn:2014xxa}. On top of these, we divide the sphere into evenly spaced grids with latitude and longitude lines separated by $10$ degrees in Fig. \ref{cs}.
The observer is placed inside the celestial sphere at an off-centered position and denoted as $O$, and the black hole is just located at the center of the celestial sphere. Photons start from the observer, travel along null geodesics, and eventually reach the celestial sphere or fall into the black hole. The images of black holes are obtained by a projection onto the observer's local frame, which will be explained in detail afterward. Here we ignore the redshift effect and focus only on the spatial distortion of these images.

In our paper, we consider that the background metric is static and spherically symmetric, which takes the form
\begin{align}
    ds^2=-f(r)dt^2+\frac{1}{f(r)}dr^2+r^2d\theta ^2+r^2 \sin^2\theta d\phi^2.
    \label{background metric}
\end{align}
Since the effective metric is non-circular and non-rotating, we assume that the effective metric takes the form
\begin{equation}
    \begin{aligned}
        ds^2_{eff}=G_{\mu \nu }dx^{\mu}dx^{\nu}=G_{tt}dt^2+G_{rr}dr^2+G_{\theta \theta}d\theta^2+G_{\phi \phi}d\phi^2+2G_{r\theta}dr d\theta .
       \label{effective metric}
    \end{aligned}
\end{equation}

Assuming that $p^{\mu}$ is the 4-momentum vector of a photon, we have the dual vector $p_{\mu}\equiv g_{\mu \nu}p^{\nu}$ in the background spacetime. However, $p^{\mu}p_{\mu}\neq0$ due to the nonlinear electrodynamic. Note that $G_{\mu\nu}p^{\mu}p^{\nu}=0$ in the effective geometry, so it is convenient to define a new dual vector
\begin{align}
    q_{\mu}\equiv G_{\mu \nu}p^{\nu}.
\end{align}
And the effective geodesic equations are given by a group of eight first-order Hamilton equations,
\begin{align}
    \frac{dq_{\mu}}{d\lambda}=-\frac{\partial \mathcal{H} }{\partial x^{\mu}},\quad \frac{dx^{\mu}}{d\lambda}=\frac{\partial \mathcal{H} }{\partial q_{\mu}},
\end{align}
where $\lambda$ is the affine parameter, and $\mathcal{H}=G^{\mu \nu}q_{\mu}q_{\nu}/2=0$ is the Hamiltonian of photons.

Due to the stationary and axisymmetry, the metric admits two Killing vectors, which correspond to two conserved quantities of geodesic motion,
\begin{equation}
    \begin{aligned}
        E =-q_t,\quad L  =q_{\phi}.
    \end{aligned}
\end{equation}
For a massless particle, $E$ and $L$ are interpreted as the energy and the angular momentum along the axis of symmetry, respectively. 

Assuming that the observer is located in the frame with the observer basis $\left\{\hat{e}_{(t)},\hat{e}_{(r)},\hat{e}_{(\theta)},\hat{e}_{(\phi)}\right\} $, which can be expanded in the coordinate basis $\{\partial t,\partial r,\partial \theta,\partial \phi\}$ as
\begin{equation}
    \begin{aligned}
        e_{(t)}       =\frac{\partial_t}{\sqrt{-g_{tt}}},\quad e_{(r)}=\frac{\partial_r}{\sqrt{g_{rr}}}, \quad e_{(\theta)}=\frac{\partial_{\theta}}{\sqrt{g_{\theta \theta}}}, \quad e_{(\phi)}    =\frac{\partial_t}{\sqrt{g_{\phi \phi }}}.           
    \end{aligned}
\end{equation}
For a photon of four-momentum $q_{\mu}$, the locally measured momentum of is
\begin{equation}
    \begin{aligned}
        p^{(t)}      & =-\hat{e}_{(t)}\cdot p=-\frac{q_t}{\sqrt{-g_{tt}}}= \frac{E}{\sqrt{-g_{tt}}},                 \\
        p^{(r)}      & =\hat{e}_{(r)}\cdot p=\frac{q_r}{\sqrt{g_{rr}}},                                              \\
        p^{(\theta)} & =\hat{e}_{(\theta)}\cdot p=\frac{q_\theta}{\sqrt{g_{\theta \theta}}},                         \\
        p^{(\phi)}   & =\hat{e}_{(\phi)}\cdot p=\frac{q_\phi}{\sqrt{g_{\phi \phi}}}= \frac{L}{\sqrt{g_{\phi \phi}}},
    \end{aligned}
\end{equation}
where we use $q_t=-E$ and $q_{\phi}=L$.
\begin{figure}[htbp]
    \begin{center}
        \includegraphics[width=15cm]{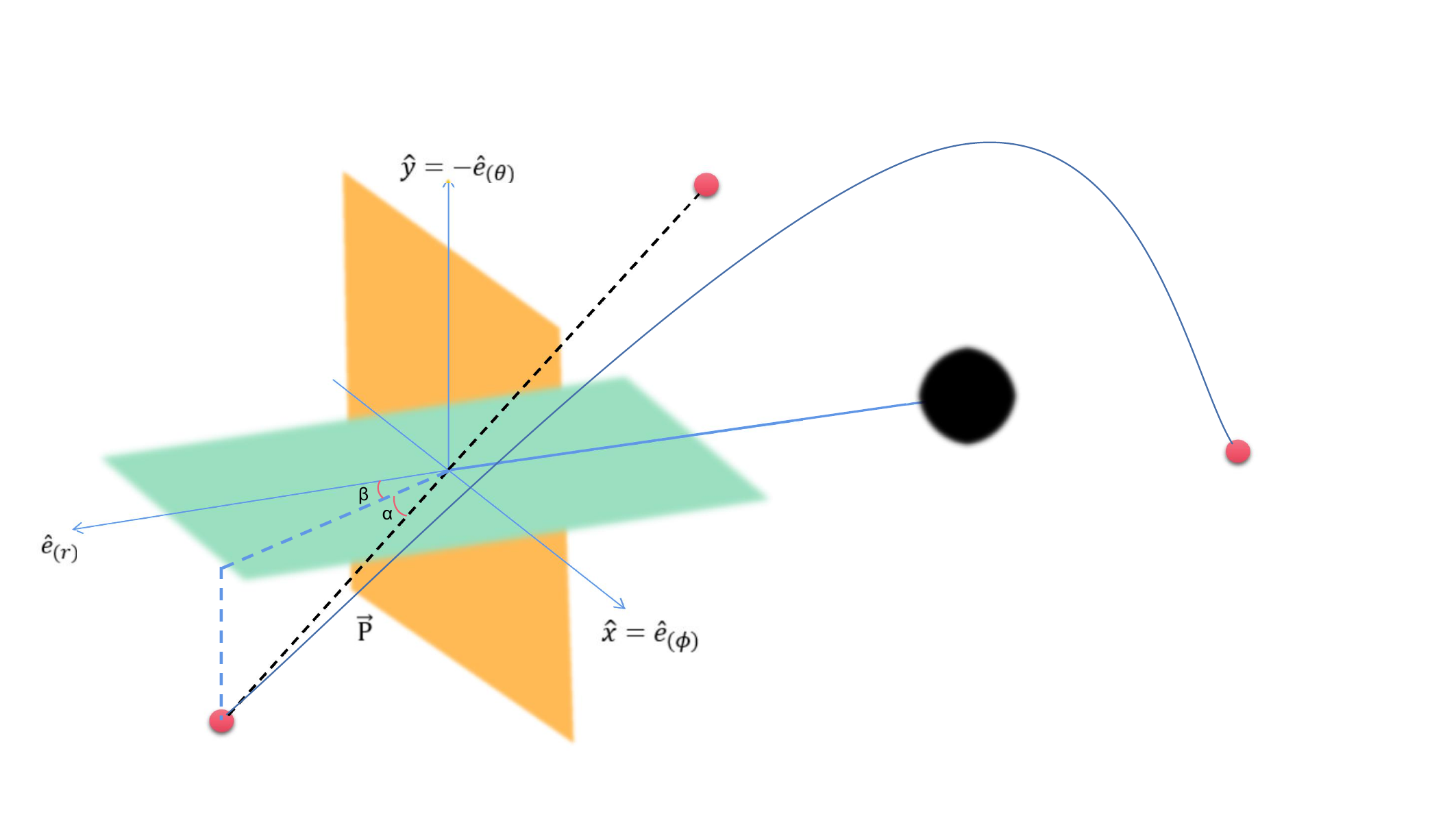}
        \caption{The vector $\vec{P}$ is the photon's 3-momentum in the observer's frame, and $(\alpha,\beta)$ are the positive observation angles. The planes associated with the angles $\alpha$ and $\beta$ are perpendicular to each other, and $\vec{P}$ is in the same plane as $\alpha$. The vectors $\hat{e}_{(\phi)}$ and $\hat{e}_{(r)}$ are also co-planar with $\beta$.}
        \label{alphabeta}
    \end{center}
\end{figure}
As in Ref. \cite{Cunha:2018acu}, we can introduce the observation angles $\alpha$ and $\beta$ in Fig. \ref{alphabeta} as
\begin{equation}
    \begin{aligned}
        \sin\alpha  =\frac{p^{(\theta)}}{|\vec{P}|}=\frac{q_{\theta}\sqrt{-g_{tt}}}{E\sqrt{g_{\phi \phi }}}, \quad
        \tan\beta   =\frac{p^{(\phi)}}{p^{(r)}}=\frac{L \sqrt{g_{rr}}}{q_r \sqrt{g_{\phi \phi }}},
    \end{aligned}
\end{equation}
where $|\vec{P}|=\sqrt{p^{(r)2}+p^{(\theta)2}+p^{(\phi)2}}=p^{(t)}$. Thus, one has
\begin{equation}
    \begin{aligned}
        p^{(r)}       =|\vec{P}|\cos\alpha \cos \beta, \quad
        p^{(\theta)}  =|\vec{P}|\sin\alpha,            \quad
        p^{(\phi)}    =|\vec{P}|\cos\alpha \sin \beta,
    \end{aligned}
\end{equation}
which can be used to express $q_t,q_r,q_{\theta},q_{\phi}$ in terms of $\alpha ,\beta$ and $|\vec{P}|$,
\begin{equation}
    \begin{aligned}
        q_t        & =-E=-\sqrt{-g_{tt}}|\vec{P}|,                          \\
        q_r        & =\sqrt{g_{rr}}|\vec{P}|\cos\alpha \cos \beta,          \\
        q_{\theta} & =\sqrt{g_{\theta \theta}}|\vec{P}|\sin\alpha,          \\
        q_{\phi}   & =L=\sqrt{g_{\phi \phi}}|\vec{P}|\cos\alpha \sin \beta.
    \end{aligned}
\end{equation}
For $p^{\mu}$, we have
\begin{equation}
    \begin{aligned}
        p^t        & =G^{tt}q_t=-G^{tt}E=-G^{tt}\sqrt{-g_{tt}}|\vec{P}|,                                                                                                                  \\
        p^r        & =G^{rr}q_r+G^{r\theta}q_{\theta}=G^{rr}\sqrt{g_{rr}}|\vec{P}|\cos\alpha \cos \beta+G^{r\theta}\sqrt{g_{\theta \theta}}|\vec{P}|\sin\alpha,                           \\
        p^{\theta} & =G^{\theta \theta}q_{\theta}+G^{\theta r}q_{r}=G^{\theta \theta}\sqrt{g_{\theta \theta}}|\vec{P}|\sin\alpha+G^{\theta r}\sqrt{g_{rr}}|\vec{P}|\cos\alpha \cos \beta, \\
        p^{\phi}   & =G^{\phi \phi}q_{\phi}=G^{\phi \phi}L=G^{\phi \phi}\sqrt{g_{\phi \phi}}|\vec{P}|\cos\alpha \sin \beta.
    \end{aligned}
\end{equation}

The Cartesian coordinates $(x,y)$ corresponding to each photon on the sky plane are functions of the observation angles $(\alpha,\beta)$. The map between them is the projection method we are considering. We introduce a well-defined distance in the curved spacetime. As in Ref. \cite{Cunha:2016bpi}, the perimetral radius or circumferential radius $\tilde{r}$ is introduced to measure the distance. It is defined as
\begin{align}
    \tilde{r}\equiv \frac{\mathcal{P}}{2\pi}=\sqrt{g_{\phi \phi}},
\end{align}
where $\mathcal{P}=\int^{2\pi}_0 \sqrt{g_{\phi \phi}}d\phi=2\pi \sqrt{g_{\phi \phi}}$ is the perimeter of a circumference at the equator $(\theta=\pi/2)$ with constant radial coordinate $r$. Assuming that the sky plane of the observer located at a distance $r=r_o$ and inclination $\theta=\theta_o$, we choose the simplest pinhole camera projection model, in which the coordinates $(x,y)$ can be presented as 
\begin{equation}
    \begin{aligned}
        x & \equiv-\tilde{r}_o\cos\alpha \sin\beta \simeq -\tilde{r}_o \beta, \\
        y & \equiv\tilde{r}_o\sin\alpha \simeq \tilde{r}_o \alpha.
    \end{aligned}
\end{equation}

The $y$-axis lies in the same plane as the black hole's rotation axis, and the black hole centers on the origin of the sky plane. The minus sign in the $x$ definition comes from the symbolic convention for $\beta$ (see Fig. \ref{alphabeta}). In addition, the vectors $\hat{e}_x$ and $\hat{e}_y$ that span the sky plane are defined as,
\begin{align}
    \hat{e}_x=\hat{e}_{(\phi)},\quad  \hat{e}_y=-\hat{e}_{(\theta)}.
\end{align}

For the geodesic equations in Eq. (\ref{geodesic equation}) and the observer's position $(r_o, \theta_o, 0)$, we can give the initial conditions at $\lambda=0$ as
\begin{equation}
    \begin{aligned}
        t(0)          & =0,r(0)=r_o,\theta(0)=\theta_o,\phi(0)=0,                                                                                   \\
        p^t(0)        & =-G^{tt}\sqrt{-g_{tt}}\vert _{r_o,\theta_o},                                                                                \\
        p^r(0)        & =[G^{rr}\sqrt{g_{rr}}\cos\alpha \cos \beta+G^{r\theta}\sqrt{g_{\theta \theta}}\sin\alpha]\vert _{r_o,\theta_o},             \\
        p^{\theta}(0) & =[G^{\theta \theta}\sqrt{g_{\theta \theta}}\sin\alpha+G^{\theta r}\sqrt{g_{rr}}\cos\alpha \cos \beta]\vert _{r_o,\theta_o}, \\
        p^{\phi}(0)   & =G^{\phi \phi}\sqrt{g_{\phi \phi}}\cos\alpha \sin \beta\vert _{r_o,\theta_o},
    \end{aligned}
\end{equation}
where we take $|\vec{P}|=1$.
The constraints are
\begin{equation}
    \begin{aligned}
        L & =q_{\phi}=G_{\phi \phi}p^{\phi}, \\
        E & =-q_t=G_{tt}p^{t},              \\
        \mathcal{H}  & =G^{\mu \nu}q_{\mu}q_{\nu}/2=0.
    \end{aligned}
\end{equation}

\end{appendices}

\bibliographystyle{unsrt}
\bibliography{egbib}

\end{document}